\documentclass[]{ceurart}

\usepackage{listings}
\usepackage{booktabs}
\usepackage{multirow}
\usepackage{amsmath}

\begin{document}

\copyrightyear{2026}
\copyrightclause{Copyright for this paper by its authors.
  Use permitted under Creative Commons License Attribution 4.0
  International (CC BY 4.0).}

\conference{Ital-IA 2026: 6th National Conference on Artificial Intelligence, organized by CINI, June 18-19, 2026, Rome, Italy}

\title{OvAi Focus: AI-based Multi-class Segmentation of Functional Ovaries
  and Adnexal Masses in Gynecological Ultrasound}

\author[1]{Niccolò Tallone}[%
email=niccolo.tallone@syndiag.ai,
]
\cormark[1]
\fnmark[1]
\author[1]{Francesca Salis}[%
email=francesca.salis@syndiag.ai,
]
\fnmark[1]
\author[1]{Pio Raffaele Fina}[%
email=pio.fina@syndiag.ai,
]
\fnmark[1]
\author[2]{Roberta Massobrio}
\author[1]{Rosilari Bellacosa Marotti}
\author[1]{Daniele Conti}
\author[2]{Luca Fuso}
\author[2]{Luca Mariani}
\author[2]{Annamaria Ferrero}
\author[3]{Alessandro Arena}
\author[4]{Stefano Cosma}
\author[5]{Dan Grisaru}
\author[7]{Angelo Lacalandra}
\author[3]{Renato Seracchioli}
\author[6]{Marianna Roccio}
\author[1]{Federica Gerace}

\cortext[1]{Corresponding author.}
\fntext[1]{These authors contributed equally.}

\address[1]{SynDiag s.r.l., Turin, Italy}
\address[2]{Academic Division of Gynecology and Obstetrics, University of Turin,
  Azienda Ospedaliera Ordine Mauriziano, Turin, Italy}
\address[3]{Division of Gynaecology and Human Reproduction Physiopathology,
  IRCCS Azienda Ospedaliero-Universitaria di Bologna, Bologna, Italy}
\address[4]{Obstetric and Gynecology Unit, Ospedale Sant'Anna, Department of
  Surgical Sciences, University of Turin, Turin, Italy}
\address[5]{Tel Aviv Sourasky Medical Center, Tel Aviv, Israel}
\address[6]{Obstetric and Gynecology Unit, Policlinico San Matteo di Pavia,
  Pavia, Italy}
\address[7]{Obstetric and Gynecology Unit, Presidio Ospedaliero Ospedale Martini,
  Turin, Italy}

\begin{abstract}
Ovarian cancer is the deadliest gynecological malignancy; accurate and
objective segmentation of adnexal masses and functional ovaries in
ultrasound (US) remains challenging due to operator variability and
morphological complexity. We present OvAi Focus (SynDiag s.r.l., Italy),
a stand-alone AI software medical device that performs multi-class semantic
segmentation of functional ovaries and adnexal masses, distinguishing
cystic from solid components. The system was trained and independently
validated on a multicenter dataset of 1{,}081 adult women from 6 centers
across Italy and Israel. Segmentation achieved DICE scores of 0.87
(complete lesion), 0.85 (cystic), 0.68 (solid), and 0.62 (functional
ovary), in line with or superior to state-of-the-art approaches across
heterogeneous acquisition settings.
\end{abstract}


\maketitle

\section{Introduction}

Ovarian cancer is the eighth most common cancer and eighth leading cause of
cancer death among women worldwide, with 324{,}603 new cases and 206{,}956
deaths estimated globally in 2022~\cite{caruso2025ovarian,iarc2022_ovary}.
The asymptomatic nature of adnexal masses in the vast majority of cases
increases the difficulty of early-stage detection. This is especially
critical for malignant lesions diagnosed at stages III--IV, where the
survival rate drops to 17--39\%~\cite{kamal2018ovarian,lu2018screening}.
Moreover, the healthy ovary exhibits high physiological and morphological
variability, thus making the discrimination of pathological findings more
complex.

Ultrasound (US) is the primary imaging modality for gynecological
evaluation; however, its diagnostic accuracy largely depends on the
operator's experience and subjective
interpretation~\cite{fischerova2015ultrasound}. The International Ovarian
Tumor Analysis (IOTA) group introduced standardized terminology, predictive
models, and risk stratification
tools~\cite{timmerman2000terms,timmerman2008simple,timmerman2016predicting,
van2014evaluating,andreotti2020rads} to overcome these limitations. Despite
the efforts, their application is still influenced by the experience level,
limiting broader adoption outside specialized
centers~\cite{meys2017simple}.

In the last decades, several AI-based approaches have targeted whole-lesion
segmentation and benign/malignant
classification~\cite{barcroft2024machine,whitney2024ai,dai2024development}:
Barcroft et al.~\cite{barcroft2024machine} achieved DICE 0.85--0.88 for
whole-mass delineation; Whitney et al.~\cite{whitney2024ai} proposed a
U-Net-based model with median DICE 0.91, without quantitative validation of
solid and cystic component separation; Dai et al.~\cite{dai2024development}
reported DICE 0.887 (internal) and 0.819 (external) for full-mass
delineation, and DICE $0.69 \pm 0.26$ for healthy ovary segmentation on
static images. Conversely, the simultaneous segmentation of solid and
cystic components alongside functional ovary delineation has not been
jointly addressed or validated in a multicenter pipeline.

This paper presents OvAi Focus, an AI software medical device that fills
this gap by providing multi-class segmentation of functional ovaries and
adnexal mass components, distinguishing cystic from solid regions.

\section{Materials \& Methods}

\subsection{OvAi Focus}

OvAi Focus is a stand-alone software medical device designed to support
clinicians in describing functional ovaries and characterizing adnexal
lesions.

\subsubsection{Segmentation Module}

\begin{figure}[ht]
  \centering
  \includegraphics[width=\linewidth]{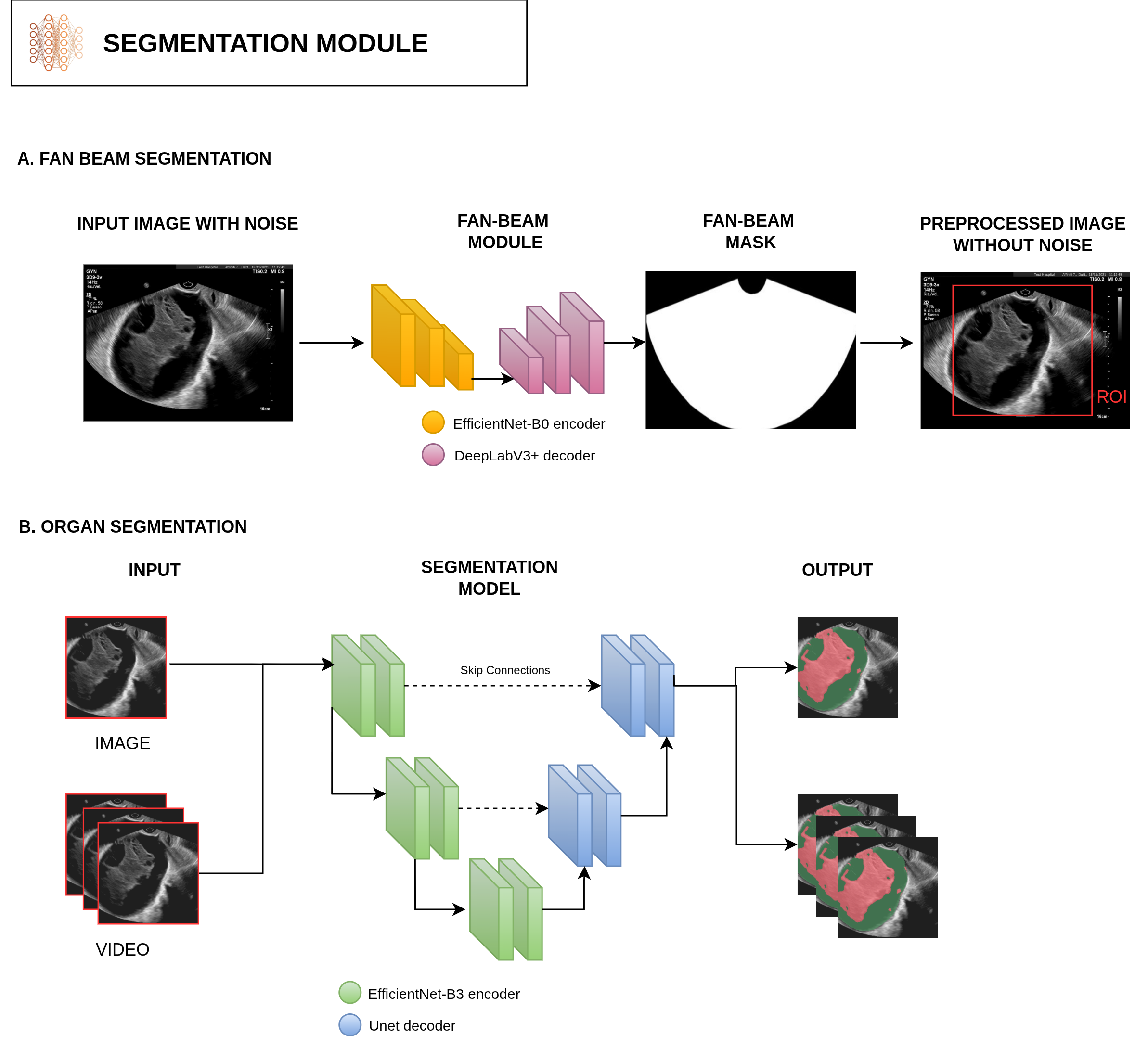}
  \caption{Segmentation module: Fan-Beam (A) and Organ Segmentation (B)
    sub-modules.}
  \label{fig2}
\end{figure}

The module operates as a cascade of two sub-modules (Figure~\ref{fig2}).

\paragraph{Fan-Beam sub-module.}
Built on DeepLabV3+~\cite{chen2018encoder} with an EfficientNet-B0
encoder~\cite{tan2019efficientnet}, this sub-module performs binary
semantic segmentation to distinguish the foreground (interior of the
fan-beam cone) from the background (the complementary area of the image).

Medical imaging data frequently contains non-anatomical information burned
into the pixel data, such as Protected Health Information (PHI) and
acquisition settings (e.g., probe orientation markers, depth, and gain
values). In US imaging, this information typically resides in the background
area outside the fan-beam cone. As demonstrated in prior
literature~\cite{vasquez2025detecting,ong2024shortcut}, deep learning models
may exploit these non-anatomical cues to form spurious correlations with the
target variable, a phenomenon known as Shortcut
Learning~\cite{geirhos2020shortcut}, thereby hindering generalization. The
Fan-Beam module addresses this by automatically detecting and isolating the
fan-beam cone, ensuring that downstream processing operates exclusively on
the masked foreground containing clinically relevant data.

\paragraph{Organ Segmentation sub-module.}
Built on a U-Net~\cite{ronneberger2015u} with an EfficientNet-B3
encoder~\cite{tan2019efficientnet}, this sub-module accepts the cropped ROI
from the Fan-Beam mask and classifies pixels into three categories:
functional ovary, cystic adnexal mass components, and solid adnexal mass
components. Predicted masks are overlaid on the US data for clinical
review.

\subsubsection{Training}

Both models were trained in a supervised setting on US data (images and
videos) paired with manually annotated ground-truth masks. Hyperparameters
(batch size, learning rate, number of epochs) were selected via k-fold
cross-validation; final production models were retrained on the full
training set and evaluated on an independent held-out test set to ensure
unbiased performance estimates.

\subsection{Dataset and Ethics}

Patients were prospectively and retrospectively enrolled from September 2019
to August 2025 across 6 clinical centers in Italy and Israel (A.O.\ Ordine
Mauriziano di Torino, A.O.U.\ Città della Salute e della Scienza Ospedale
Sant'Anna, Presidio Ospedaliero Ospedale Martini, IRCCS AOU Policlinico
Sant'Orsola Malpighi, Policlinico San Matteo di Pavia, and Tel Aviv
Sourasky Medical Center), following IRB approval at each site and the
Declaration of Helsinki. A subset of 1{,}081
adult women was selected (mean age $52.45 \pm 16.40$ years), including cases
with functional ovaries only, adnexal masses only, and both. Data were
acquired on multiple commercial sonograph platforms (Philips, GE, Samsung)
using transvaginal and/or transabdominal probes.

The segmentation dataset comprised 1{,}004 clinical cases: training set of
745 cases (499 adnexal masses only, 135 functional ovaries only, 111 with
both structures) for a total of 1{,}801 items (804 videos, 997 images;
1{,}180 B-mode, 621 Doppler); independent test set of 259 cases (175
adnexal masses only, 47 functional ovaries only, 37 with both structures)
for a total of 676 items (364 videos, 312 images; 420 B-mode, 256 Doppler).

\subsection{Evaluation}

Segmentation performance was evaluated by comparing predicted masks against
manually annotated ground-truth masks. A clinical team trained and supervised
by expert gynecologists labeled ground-truth masks of functional ovaries,
cystic components, and solid components using the Labelbox platform
(Labelbox Inc., USA); each mask was individually reviewed by an expert
medical doctor. Performance metrics (IoU, DICE, Precision, Recall) were
aggregated via micro-averaging~\cite{reinke2024understanding}, which pools
all pixels across the dataset and weights contributions proportionally to
structure size, thereby reflecting expected performance over the full dataset
rather than overemphasizing rare or atypical cases.

\section{Results}

Table~\ref{tab:seg} reports segmentation performance on the independent test
set. The model achieved DICE 0.87 for the complete lesion, 0.85 for cystic
components, 0.68 for solid components, and 0.62 for functional ovary. All
metrics follow a consistent trend across categories, confirming the
robustness of the evaluation.

\begin{table}[ht]
  \caption{Segmentation performance on the independent test set.}
  \label{tab:seg}
  \centering
  \begin{tabular}{lcccc}
    \toprule
    \textbf{Metric} & \textbf{Lesion solid} & \textbf{Lesion cystic}
      & \textbf{Complete lesion} & \textbf{Functional ovary} \\
    \midrule
    IoU       & 0.51 & 0.74 & 0.77 & 0.45 \\
    DICE      & 0.68 & 0.85 & 0.87 & 0.62 \\
    Precision & 0.75 & 0.85 & 0.90 & 0.63 \\
    Recall    & 0.62 & 0.85 & 0.85 & 0.61 \\
    \bottomrule
  \end{tabular}
\end{table}

\section{Discussion}

Segmentation results are in line with or superior to comparable
state-of-the-art approaches. The model achieved its best results for cystic
components (DICE 0.85, IoU 0.74), attributable to the homogeneous content
and clearer US boundaries characteristic of cystic regions. Solid components
were more challenging (DICE 0.68, IoU 0.51), reflecting their heterogeneous
echogenic patterns and less well-defined borders. To our knowledge, no
reference values exist in the literature for solid and cystic component
segmentation separately, highlighting the novelty of this contribution.

Complete lesion DICE (0.87) is comparable to Dai et
al.~\cite{dai2024development} (0.89 internal, 0.82 external), Barcroft et
al.~\cite{barcroft2024machine} (0.85), and Whitney et
al.~\cite{whitney2024ai} (0.91 median). Functional ovary DICE (0.62) falls
within the uncertainty range of Dai et al.\ ($0.69 \pm 0.26$), despite our
dataset including both images and videos and a higher prevalence of mass
cases (83\% of the cohort) relative to functional ovaries (30\%), and
despite the lack of sharp anatomical boundaries and high inter-patient
morphological variability of the ovary. All metrics exhibit a
consistent trend across categories, confirming that relative segmentation
performance is independent of the metric adopted and that observed
differences reflect the intrinsic US characteristics of each structure.

\section{Conclusion}

OvAi Focus is the first validated AI software medical device to
simultaneously segment functional ovaries and adnexal mass components
(cystic and solid) in gynecological US. Validated on an independent
multicenter dataset of 1{,}081 patients, it achieves state-of-the-art
whole-lesion segmentation and extends multi-class delineation to solid and
cystic components separately --- a task not previously validated in the
literature. Strengths include multicenter design, independent test sets,
strict data-leakage prevention, and heterogeneous acquisition settings.
Future work will expand dataset diversity, integrate the device into
prospective clinical workflows, and extend validation to additional
segmentation targets.

\section*{Declaration on Generative AI}

During the preparation of this work, the authors used Gemini (Google) in order to: Grammar and spelling check. After using this tool, the authors reviewed and edited the content as needed and take full responsibility for the publication's content.

\bibliography{bibliography}

\end{document}